\documentclass[%
 reprint,
 floatfix,
 amsmath,amssymb,
 aps,
]{revtex4-2}
\usepackage{graphicx}
\usepackage{dcolumn}
\usepackage{bm}
\usepackage{caption}
\usepackage{physics}
\usepackage{hyperref}
\usepackage{float}
\usepackage{booktabs}
\begin{document}
\preprint{APS/123-QED}
\title{Developing a platform for linear mechanical quantum computing}
\author{Hong Qiao$^{1}$}
\author{\'Etienne Dumur$^{1,2}$}
\altaffiliation[Present Address: ]{Univ. Grenoble Alpes, CEA, Grenoble INP, IRIG, PHELIQS, 38000 Grenoble, France}
\author{Gustav Andersson$^{1}$}
\author{Haoxiong Yan$^{1}$}
\author{Ming-Han Chou$^{1,3}$}
\author{Joel Grebel$^{1}$}
\author{Christopher R. Conner$^{1}$}
\author{Yash J. Joshi$^{1}$}
\author{Jacob M. Miller$^{1,3}$}
\author{Rhys G. Povey$^{1,3}$}
\author{Xuntao Wu$^{1}$}
\author{Andrew N. Cleland$^{1,2}$}
\email{Corresponding author: anc@uchicago.edu.}

\affiliation{$^{1}$Pritzker School of Molecular Engineering, University of Chicago, Chicago IL 60637, USA}
\affiliation{$^{2}$Center for Molecular Engineering and Material Science Division, Argonne National Laboratory, Lemont IL 60439, USA}
\affiliation{$^{3}$Department of Physics, University of Chicago, Chicago IL 60637, USA}


\begin{abstract}
Linear optical quantum computing provides a desirable approach to quantum computing, with a short list of required elements. The similarity between photons and phonons points to the interesting potential for linear mechanical quantum computing (LMQC), using phonons in place of photons. While single-phonon sources and detectors have been demonstrated, a phononic beamsplitter element remains an outstanding requirement. Here we demonstrate such an element, using two superconducting qubits to fully characterize a beamsplitter with single phonons. We further use the beamsplitter to demonstrate two-phonon interference, a requirement for two-qubit gates, completing the toolbox needed for LMQC.  This advance brings linear quantum computing to a fully solid-state system, along with straightforward conversion between itinerant phonons and superconducting qubits.  
\end{abstract}

\maketitle

Linear optical quantum computing (LOQC) presents a scalable approach to quantum computing that relies only on relatively simple optical elements such as beamsplitters, phase shifters, and single phonon sources and detectors \cite{knill2001}. The similarity between photons and phonons poses the question as to whether linear \emph{mechanical} quantum computing (LMQC) might be achieved using phonons.

Prior experiments with phonons in solid systems have included the quantum control of mechanical motion \cite{OConnell2010, Chu2017, Satzinger2018}, entanglement between macroscopic mechanical objects \cite{Palomaki2013, Riedinger2018, OckeloenKorppi2018, Wollack2022}, coupling between surface acoustic waves and qubits \cite{Gustafsson2014, Manenti2017, Noguchi2017, Moores2018, Bolgar2018}, the deterministic emission and detection of individual SAW phonons \cite{Bienfait2019, Dumur2021}, and the transmission of quantum information \cite{Bienfait2019, Bienfait2020, Dumur2021, Zivari2022, Zivari2022a}, among other demonstrations \cite{Delic2020, Shao2022}. 

Here we explore the potential for linear quantum computing, extending this prior work by demonstrating a phonon beamsplitter for surface acoustic wave (SAW) phonons, first showing that the beamsplitter deterministically converts a single incident phonon to a superposition output state, with one phonon in either of the two output channels. This is a phase-coherent process, which we further exploit to demonstrate a single-phonon interferometer, using qubits to control the phonon phase.  We further explore two-phonon interference via the Hong-Ou-Mandel (HOM) effect \cite{Hong1987}, central to a controlled-phase gate in LOQC \cite{knill2001, Kok2007}, using two SAW phonons whose simultaneous arrival suppresses the output of coincident phonons in the two output channels, in favor of a superposed two-phonon-per-channel output, with a suppression visibility of $0.910 \pm 0.013$.

These results show that we now have the basic toolset for linear mechanical quantum computing, perhaps surprising given that in our system, a single phonon represents the collective motion of a large number ($\sim 10^{15}$) of atoms. 
\begin{figure*}[t!]
\begin{center}
	\includegraphics[width=1\textwidth]{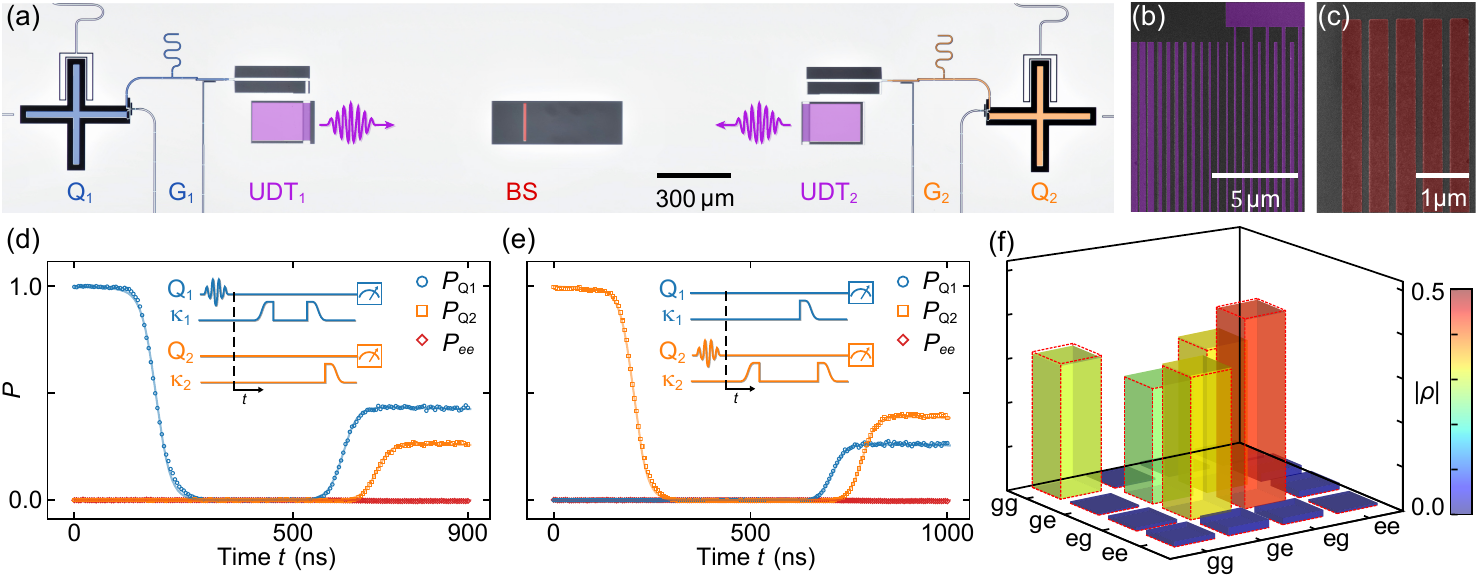}
	\caption{
    \label{fig1}
    {\bf Device description and characterization.}
    (a) False-color optical micrograph of two superconducting qubits ($\text{Q}_1$ and $\text{Q}_2$) coupled to unidirectional transducers ($\text{UDT}_1$ and $\text{UDT}_2$) via tunable couplers ($\text{G}_1$ and $\text{G}_2$), on either side of a phonon beamsplitter (BS). Qubits and couplers are fabricated on a sapphire die, with UDTs and BS on a separate, smaller lithium niobate die; micrograph taken from the backside of the flip-chip assembled device. (b) False-color scanning electron micrograph of transducer-mirror UDT combination, and (c) top-left corner of BS. (d), (e) A single phonon is sent from $\text{Q}_1$ ($\text{Q}_2$) to the BS, ``split'' by the beamsplitter and the output subsequently captured by each qubit. Measurements of both qubits are made at time $t$, yielding the excited state populations $P_\text{Q1}$ (blue circles),  $P_\text{Q2}$ (orange squares), and the joint excitation probability $P_{ee}$ (red diamonds). Solid color lines are numerical simulations (see Supplemental Material). Insets: Control pulse sequences. (f) After the qubits capture the phonon emitted in (D), we perform two-qubit state tomography yielding the two-qubit density matrix $\rho$, with Bell state fidelity $\mathcal{F} = 0.816\pm0.004$. The tomography measurement is repeated 10 times and the density matrix is reconstructed in each repetition and constrained to be Hermitian. All uncertainties are one standard deviation. Bars represent measured $|\rho|$, and red dashed frames are simulated values.}
\end{center}
\end{figure*}

Our device comprises two superconducting Xmon qubits $\text{Q}_1$ and $\text{Q}_2$ \cite{Koch2007,Barends2013}, coupled via two tunable couplers $\text{G}_1$ and $\text{G}_2$ \cite{Chen2014} to two unidirectional interdigitated transducers $\text{UDT}_1$ and $\text{UDT}_2$ \cite{Dumur2021}. These are linked by a 2 mm-long SAW phonon channel, interrupted by a phonon beamsplitter BS (Fig.~\ref{fig1}(a)-(c)). The beamsplitter, comprising a set of 16 parallel metal fingers, is designed to reflect approximately half of the acoustic signal, while transmitting the remainder. Details for the qubit, UDT, and BS designs appear in the Supplemental Material.  The BS is intentionally positioned 150 nm closer to $\text{UDT}_1$ than $\text{UDT}_2$, resulting in slightly different travel times for phonon wavepackets emitted simultaneously from the two UDTs. The variable coupler between each qubit and its associated UDT allows us to shape the phonon emission, with typical emission times ranging from 14 ns (maximum coupling) to more than 10 $\mu$s  (minimum, ``off'' coupling), which we characterize via the couplers' time-dependent emission rates $\kappa_{1,2}(t)$ (see Fig.~S2) \cite{Bienfait2019}. 

The qubits, variable couplers, and their associated control and readout lines are fabricated on a sapphire substrate, while the acoustic elements (UDTs and BS) are fabricated on a separate lithium niobate substrate. Following fabrication, the two dies are aligned and attached to one another using a flip-chip assembly \cite{Satzinger2019}. The device is operated in a dilution refrigerator with a base temperature of about 10 mK. 

We first characterize the system by measuring the response to single phonons. One qubit (either Q$_1$ or Q$_2$) is excited to its $|e\rangle$ state and a phonon is emitted, where we shape the emission to have a hyperbolic secant waveform, $\phi_{1,2}(t) \propto \sech(t/2 \sigma_{1,2})$, via the calibrated time-dependent modulation of the qubit's variable coupling rate $\kappa_{1,2}(t)$ \cite{Bienfait2019}; the characteristic wavepacket width is $\sigma_{1,2}=17.9$ ns. The phonon released from Q$_1$ (Q$_2$) interacts with the beamsplitter, ideally resulting in the output state $\left (i|10\rangle + |01\rangle\right )/\sqrt{2}$ ($\left (|10\rangle + i|01\rangle\right )/\sqrt{2}$); see also Fig.~S1(b),(c). We use the notation $\ket{\text{ph}_{1} \text{ph}_{2}}$, where $\text{ph}_{1,2}$ denotes the phonon number in the output channel directed towards $\text{Q}_{1,2}$, respectively.  The beamsplitter output is then captured by both qubits, via a calibrated time-dependent variation of each qubit's coupling rate. In Fig.~\ref{fig1}(d), (e), we display the excited state probability $P_e(t)$ for each qubit and their joint excitation probability $P_{ee}(t)$ as a function of measurement time $t$. The joint excitation probability $P_{ee}(t)$ remains very small, consistent with the expectation of no joint qubit excitation in this measurement, as each experiment involves only one phonon at a time. From these data, we extract a beamsplitter reflectivity $\eta=0.61$ and effective itinerant phonon lifetime $\tau_{ph} = 1.3~\mu$s, in agreement with a separate characterization of the BS design (see Fig.~S1(d)) and previous measurements of phonon propagation loss in similar systems \cite{Bienfait2019, Dumur2021}. We note the effective lifetime includes loss from phonon scattering during transits between the UDTs and the BS, from the UDT and BS elements themselves, as well as any losses during qubit release or capture. 

We perform two-qubit state tomography for the final joint qubit state generated in Fig.~\ref{fig1}(d), displaying the absolute value of the density matrix $\rho$ in Fig.~\ref{fig1}(f). We find a Bell state fidelity $\mathcal{F} = \sqrt{\Tr(\rho_{\rm Bell}\cdot|\rho|)} = 0.816\pm0.004$ to the ideal Bell state $\rho_{\rm Bell}$, indicating the phonon beamsplitter maintains quantum coherence. Decay in the principal density matrix elements is consistent with phonon propagation and scattering loss, included via the effective phonon lifetime in the simulations that generate the dashed frames.  

\begin{figure}[t!]
\begin{center}
	\includegraphics[width=0.48\textwidth]{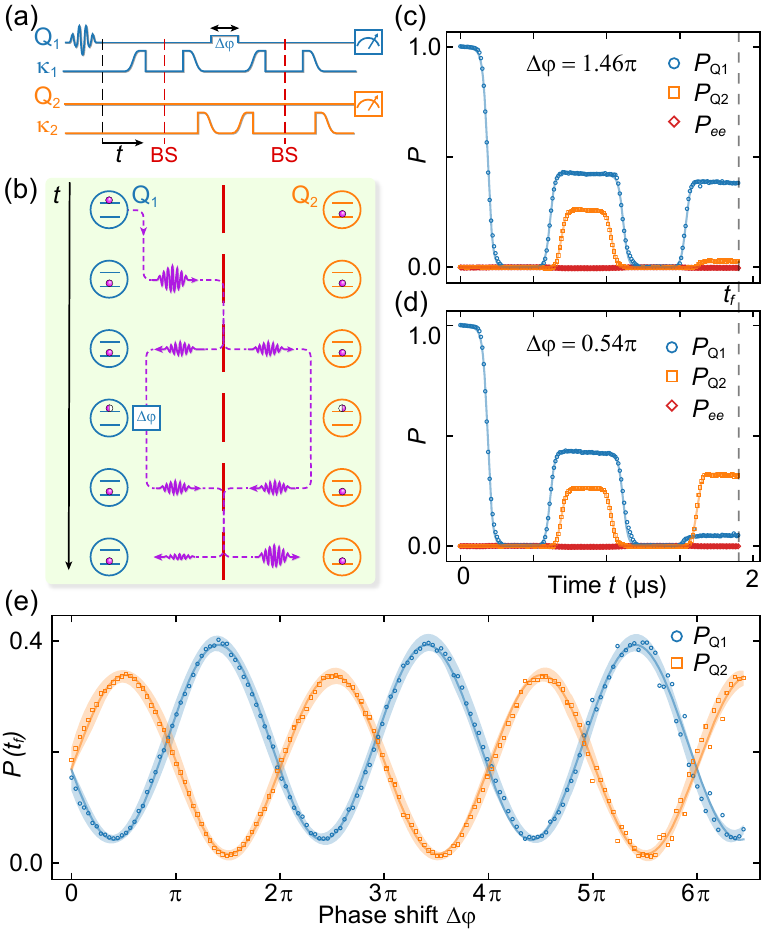}
	\caption{
    \label{fig4}
    {\bf Single phonon interferometry.}
    (a) Control pulse sequence. (b) Schematic representation: A single phonon is emitted by $\text{Q}_1$, split by the BS, and captured by both qubits, a process identical to the Bell state generation shown in Fig.~\ref{fig1}(d). A brief frequency change of Q$_1$ adds a phase shift $\Delta \varphi$, followed by a phonon release from each qubit, timed to give coincident signals at the BS and thus generating interference. (c), (d) Deterministic routing of the BS interference output to either $\text{Q}_1$ or $\text{Q}_2$, respectively, for phase shifts $\Delta \varphi = 1.46\pi$ or $0.54\pi$. Solid blue, orange and red lines are simulations. (e) Excitation probability for $\text{Q}_1$ and $\text{Q}_2$ as a function of the control phase $\Delta \varphi$, showing an interference pattern with visibility $\mathcal{V}_\text{Q1} = 0.806 \pm 0.004$ and $\mathcal{V}_\text{Q2} = 0.910 \pm 0.005$. The relative phase $\Delta \varphi$ is varied by changing $\text{Q}_1$'s frequency for a variable length of time, as illustrated in the pulse sequence. Light blue and orange shaded areas denote one standard deviation calculated from 34 repeated scans. Solid blue and orange lines are cosine fits.}
\end{center}
\end{figure}

We further demonstrate the coherence of the BS element by performing a Mach-Zehnder--like interference experiment using a single itinerant phonon. In Fig.~\ref{fig4}(a) we display the control pulse sequence, in Fig.~\ref{fig4}(b) a schematic representation of the experiment, and in Fig.~\ref{fig4}(c-e) the results from this experiment. A single phonon is emitted from Q$_1$ with the same waveform as in Fig.~\ref{fig1}(d), and the beam-split phonon is captured by Q$_1$ and Q$_2$. The phase of Q$_1$ relative to Q$_2$ is then changed by $\Delta \varphi$, and the excitations re-emitted via a timed release from each qubit, resulting in a zero-delay interference at the BS. The phase-dependent interference allows control of which channel receives the output phonon, shown by plotting the excited state probabilities $P_{\rm Q1,2}$, as well as the joint excitation probability $P_{ee}$, for two choices of phase $\Delta \varphi = 1.46\pi$ and $0.54\pi$ in Fig.~\ref{fig4}(c) and (d); these choices of phase result in routing the maximal output phonon population to Q$_1$ or Q$_2$, respectively. In Fig.~\ref{fig4}(e) we display the resulting high-visibility interference fringes for the final excitation probabilities $P_{\rm Q1,2}(t_f)$ as a function of $\Delta \varphi$. The interference fringe visibility for Q$_1$ is $\mathcal{V}_\text{Q1}=0.806\pm0.004$ and for Q$_2$, $\mathcal{V}_\text{Q2}=0.910\pm0.005$, where the visibilities are defined as $\mathcal{V}_\text{Q1,2} = (P_\text{Q1,2,max}-P_\text{Q1,2,min})/(P_\text{Q1,2,max}+P_\text{Q1,2,min})$. We note that there is a slight misalignment in the interference pattern for $P_\text{Q1}$ compared to $P_\text{Q2}$, so that e.g. the phases for Fig.~\ref{fig4}(c) and (d) do not differ by exactly $\pi$. This is possibly because the phase difference between reflected and transmitted phonons is not exactly $\pi/2$ (see Fig.~S1(c) and discussion in \cite{Uppu2016}). 

\begin{figure}[t!]
\begin{center}
\includegraphics[width=0.45\textwidth]{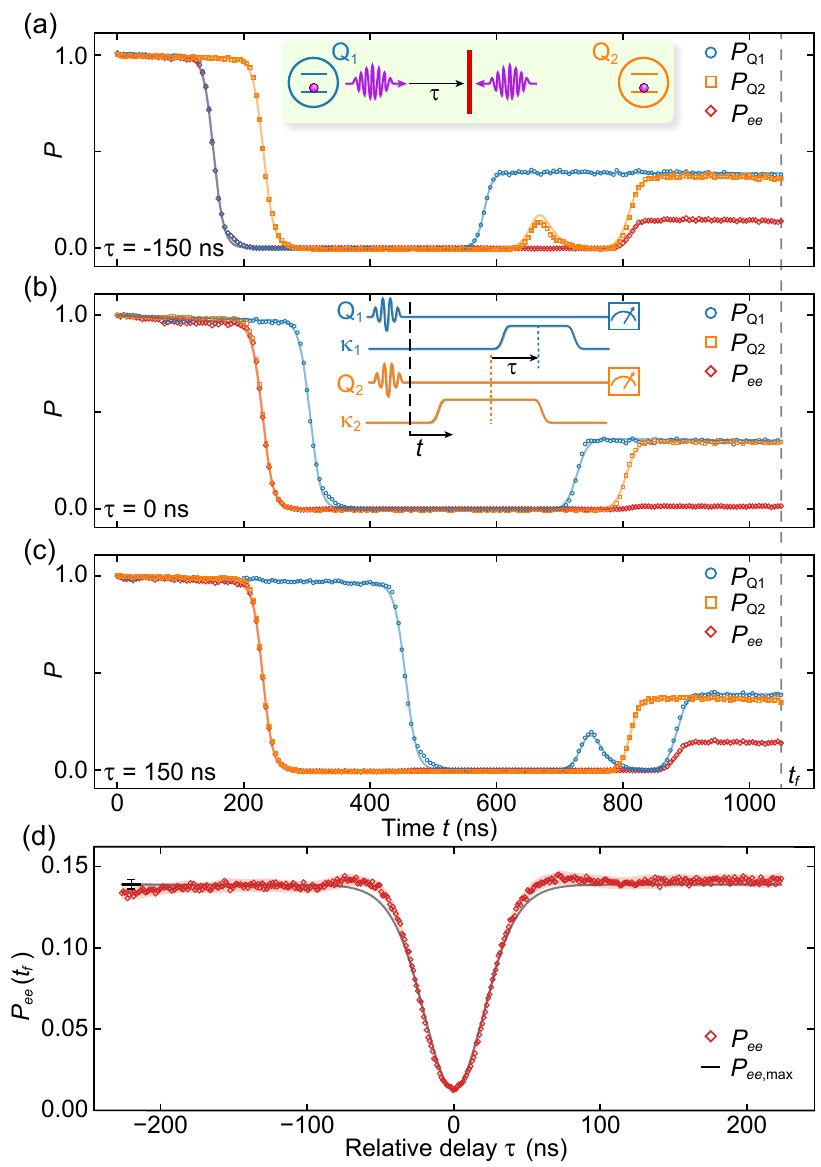}
\caption{
\label{fig2}
{\bf Two-phonon interference.}
Each qubit is initialized to its excited state $|e\rangle$, and the qubit coupling rates $\kappa_{1,2}$ controlled to emit a single phonon from each qubit, timed so the centers of the phonon wavepackets reach the center of the beamsplitter with a relative delay of (a) $\tau=-150$ ns, (b) 0 ns, and (c) 150~\text{ns}, with extracted phonon wavepacket widths of $\sigma_1=8.4~\text{ns}$ (Q$_1$) and $\sigma_2=8.3~\text{ns}$ (Q$_2$). The output waveforms from the beamsplitter are subsequently captured by the qubits via a tuned variation of their coupling rates, and simultaneous measurements of the qubits made at time $t$. Solid lines are numerical simulations (see Supplemental Material).  A schematic of the process appears inset to panel (a), and a pulse sequence inset to panel (b). The coupling remains on between release and catch. Blue and orange dashed lines in the pulse sequence denote phonon arrival times at the BS. (d) Two-qubit joint excitation probability $P_{ee}$ measured at time $t_f$ as a function of relative delay $\tau$, showing a pronounced Hong-Ou-Mandel dip for coincident phonons. The visibility of the dip is $\mathcal{V} = 0.910 \pm 0.013$, with $P_{ee,\text{max}} = 0.139 \pm 0.003$ marked in the upper-left corner (black error bar). Gray line is $\alpha \, P_{11}$, calculated from Eq.~(\ref{eq1}) with parameters extracted from panels (a-c) (see main text). Light red shaded area denotes one standard deviation, calculated from ten repeated scans for each delay $\tau$.}
\end{center}
\end{figure}

We next study two-phonon interference, shown in Fig.~\ref{fig2}, with the schematic process shown inset to Fig.~\ref{fig2}(a). One phonon is emitted by each qubit, timed so that the phonons arrive at the beamsplitter with a relative delay $\tau$; the beamsplitter output is then captured by the two qubits. The probability $P_{11}$ to have coincident single phonons in the outputs of a lossless beamsplitter with reflectivity $\eta$ is
\begin{equation}\label{eq1}
    P_{11}(\tau) = 1 - 2\eta + 2\eta^2 + (2\eta^2 - 2\eta) \left [\int \phi_1(t-\tau) \phi_2(t) dt \right ]^2
\end{equation}
(see Supplemental Material). For large delays, the integral is zero, so $P_{11}(\tau \gg \sigma_{1,2}) \rightarrow 1 - 2 \eta + 2 \eta^2 = 0.524$, using the measured beamsplitter reflectivity $\eta$. For zero relative delay and identical time-matched waveforms, the integral is unity, giving $P_{11}(0) = (1 - 2 \eta)^2 = 0.048$. The coincident probability $P_{11}$ is thus strongly suppressed for zero delay compared to large delays, with a theoretical visibility $\mathcal{V}_{\rm th} \equiv [P_{11}(\tau \gg \sigma_{1,2})-P_{11}(0)]/P_{11}(\tau \gg \sigma_{1,2}) = 0.908$. 

We cannot directly measure the coincident phonon probability $P_{11}$, but the qubit joint excitation probability $P_{ee}$ serves as a proxy: The probability for both qubits to be excited is closely related to the probability of having a phonon in each output channel. The two probabilities are not in general simply related, due in part to the non-zero probability of the two-phonon state in each output channel, as well as the different phonon loss and imperfect phonon capture by the qubits in the two channels. However, we find experimentally that $P_{ee}$ and $P_{11}$ are closely proportional, with $P_{ee} \approx \alpha P_{11}$ with an empirical scale factor $\alpha = 0.265$. The experimental $P_{ee}$ is also in good agreement with simulations for large and for zero delay (see Fig.~\ref{fig2}(a-c)).
 
 We show the experimental pulse sequence inset to Fig.~\ref{fig2}(b): The qubits are calibrated to emit phonons at 3.925 GHz, and the variable couplers are tuned to release phonons with a hyperbolic secant shape, with fit wavepacket widths $\sigma_{1,2}=8.4$ ns and 8.3 ns, respectively, timed so the phonons arrive at the beamsplitter with relative time delay $\tau$. After emission, the qubits are in their ground states $|g\rangle$. The phonons output from the beamsplitter are  captured by the two qubits, using the reverse process to emission \cite{Bienfait2019, Dumur2021}, and the two qubits measured simultaneously.  The capture process is timed so that each qubit catches its own beam-split phonon, with an efficiency very similar to the single phonon experiment in Fig.~\ref{fig1}(d) and (e). 

In Fig.~\ref{fig2}(a-c) we show the excited state probabilities for the two qubits for three different relative delays $\tau=-150$ ns, $\tau=0$ ns and $\tau=150$ ns, including the full emission and capture process. In Fig.~\ref{fig2}(d) we show the corresponding joint excitation probability $P_{ee}$ as a function of delay time $\tau$. When the delay $\tau \gg \sigma_{1,2}$, phonons pass independently through the beamsplitter, yielding a joint qubit excitation probability $P_{ee,\text{max}} = P_{\text{Q}1} \times P_{\text{Q}2} = 0.139 \pm 0.003$. However, when the two phonons arrive at the beamsplitter with zero delay $\tau$, two-phonon interference suppresses the coincident phonon output state $\ket{11}_{ph}$, and similarly suppresses $P_{ee}$, with a minimum at zero delay of $P_{ee,\text{min}} = 0.0125 \pm 0.0018$. The visibility of the dip in $P_{ee}$, $\mathcal{V}_{\rm exp} \equiv (P_{ee,\text{max}}-P_{ee,\text{min}})/P_{ee,\text{max}} = 0.910 \pm 0.013$, agrees well with the visibility calculated for $P_{11}$. We note in the ``catch'' portion of this experiment, the qubits do not capture the entire phonon signal; this can be seen in Fig.~\ref{fig2}(a), where the transmitted portion of the phonon from $\text{Q}_1$ briefly excites $\text{Q}_2$ at $t \sim 670$ ns. As $\text{Q}_2$'s coupler is left on, this excitation is subsequently re-emitted. As in this process, the two qubits are detecting the same phonon emitted by Q$_1$, only one qubit can be excited at a time, so $P_{ee}$ remains small in this portion of the measurement. There is a similar behavior at $t \sim 750$ ns in Fig.~\ref{fig2}(c), involving the phonon emitted by Q$_2$. For zero delay, the two-phonon signal is only partially captured by the receiving qubits, as each qubit can only catch a single phonon; this is similar to the response of non-number-resolving photon detectors (see Fig.~S3, showing a variation of this measurement process). 

\begin{figure}[ht!]
\begin{center}
	\includegraphics[width=0.48\textwidth]{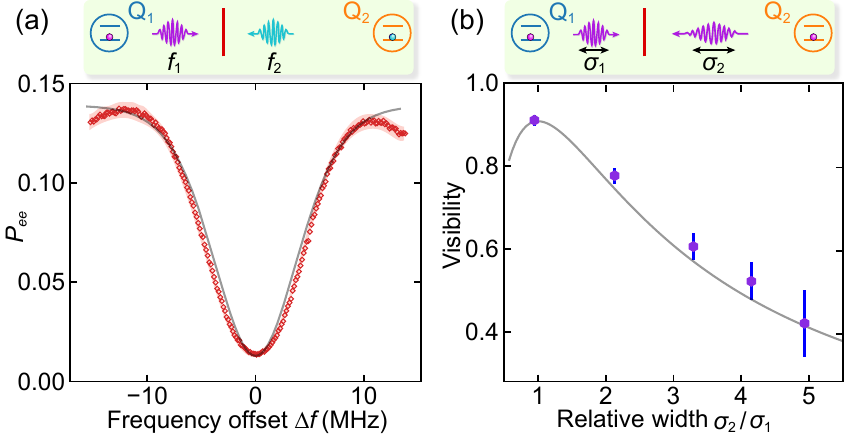}
	\caption{
    \label{fig3}
    {\bf Frequency and wavepacket dependence of two-phonon interference.}
(a) Top: Schematic of phonons released with different frequencies $f_{1,2}$. Main panel: Setting the frequency of $\text{Q}_1$ to $f_{1}$ = 3.925 GHz, the relative frequency of $\text{Q}_2$ is varied (horizontal axis) with zero relative delay $\tau$ while monitoring the joint excitation probability $P_{ee}$ following phonon capture (vertical axis). Wavepacket widths are fixed at $\sigma_{1,2}=16.0~\text{ns}$, 16.7 ns. A high visibility dip $\mathcal{V} = 0.899 \pm 0.013$ is observed at zero frequency difference. Gray line is $\alpha \, P_{11}$ from frequency-domain theoretical model in Supplemental Material. (b) Top: Schematic of phonons released with different wavepacket widths $\sigma_{1,2}$. Main panel: We fix both phonon center frequencies at 3.925 GHz as well as the width $\sigma_1 = 8.8$ ns for Q$_1$, while varying that for Q$_2$ over the values $\sigma_2 = 18.7$ ns, 28.9 ns, 36.4 ns and 43.3 ns, and perform two-phonon interference experiment. We plot the relative phonon wavepacket width $\sigma_2/\sigma_1$ (horizontal axis) and show the resulting HOM dip visibility (vertical axis). The measured visibility is maximal when the wavepackets have equal widths, then falls with increasing relative width. First data point is from Fig.~\ref{fig2}(d), and the other data are from Fig.~S4. Gray line is calculated from Eq.~(\ref{eq1}).}
\end{center}
\end{figure}

Finally, we study the indistinguishability of the two phonons by varying their relative center frequencies and wavepacket widths. While fixing the phonon wavepacket widths $\sigma_{1,2}=16.0$ ns and 16.7 ns, setting the delay $\tau$ to zero, and fixing $\text{Q}_1$'s frequency to $f_1=3.925$ GHz, we systematically vary Q$_2$'s frequency $f_2$ using the qubit flux control. In Fig.~\ref{fig3}(a), we display the resulting measured joint excitation probability $P_{ee}$ as a function of relative frequency detuning $\Delta f = f_1-f_2$. We observe a clear dip in $P_{ee}$ with frequency $\Delta f$, with visibility $\mathcal{V} = 0.899 \pm 0.013$. The full width at half maximum of the dip is 9.5 MHz, close to the bandwidth of the phonon wavepackets. A similar experiment has been done with optical photons \cite{Imany2018}, with similar results. Next, to study the waveform width dependence, we vary the wavepacket width $\sigma_2$ of $\text{Q}_2$ while fixing that of $\text{Q}_1$ to $\sigma_1=8.8$ ns. With a similar pulse sequence to that shown in Fig.~\ref{fig2}(b), we perform a two-phonon interference experiment, with results displayed in Fig. \ref{fig3}(b). The observed variation in visibility is in excellent agreement with the prediction of Eq.~(\ref{eq1}). We note that this experiment can also be performed while arbitrarily varying the functional dependence of the phonon waveform; we show an example of this in Fig.~S5.

In conclusion, we demonstrate a quantum-coherent phonon beamsplitter that deterministically generates single-phonon superpositions, suitable for interferometry as shown by a Mach-Zehnder--like experiment. We further demonstrate two-particle interference, with a visibility of $\mathcal{V} = 0.910 \pm 0.013$. As part of this second experiment, we systematically study the indistinguishability of the interfering phonons in terms of their arrival times, center frequencies and waveforms.  One limitation in this experiment is that we cannot directly detect a two-phonon state; in Fig.~S3 we explore how this limitation can be ameliorated.  An outstanding question is whether phonon loss can be made sufficiently small to enable scaling to useful computational systems.

These experiments demonstrate a phonon testbed that can generate arbitrary phonon pulse shapes with time-frequency multiplexing capability. The easy integration with superconducting qubits, which enable the on-demand generation, detection and phase control of individual phonons, should support the development of local phononic communication networks \cite{Habraken2012, Fu2019, Taylor2022, Kuzyk2018, Lemonde2018, Chen2022}, in addition to linear mechanical quantum computing \cite{knill2001, Kok2007}. \\
~\\
H.Q. designed and fabricated the devices, performed the measurements and analysed the data. \'E.D. contributed to UDT design and measurement techniques. G.A., H.Y., M.H.C. and J.G. provided suggestions with measurement and data analysis. A.N.C. advised on all efforts. All authors contributed to discussions and production of the manuscript.

\section*{Acknowledgments}
We thank Audrey Bienfait and Peter Duda for helpful discussions. Devices and experiments were supported by the Air Force Office of Scientific Research and the Army Research Laboratory. Results are in part based on work supported by the U.S. Department of Energy Office of Science National Quantum Information Science Research Centers. \'E.D. was supported by LDRD funds from Argonne National Laboratory. This work was partially supported by UChicago's MRSEC (NSF award DMR-2011854) and by the NSF QLCI for HQAN (NSF Award 2016136). We made use of the Pritzker Nanofabrication Facility, which receives support from SHyNE, a node of the National Science Foundation's National Nanotechnology Coordinated Infrastructure (NSF Grant No. NNCI ECCS-2025633).



\bibliography{apssamp}


\clearpage

\onecolumngrid
\begin{center}
\textbf{\Large{Supplemental Material}}
\end{center}


\subsection{Theoretical modelling of two-phonon interference}
Using a similar procedure to \cite{Branczyk2017}, we derive the two-phonon coincidence probability $P_{11}$ in Eq.~(\ref{eq1}). We assume two arbitrary phonon wavepackets with the same center frequency, characterized by the temporal wavefunctions $\phi_{1,2}(t)$, normalized such that $\int dt |\phi_{1,2}(t)|^2 = 1$. The phonon creation operators $\vb{a}^\dag$ ($\vb{b}^\dag$) for phonons to the left (right) of the BS can be written as $\vb{a}^{\dag}_{\phi_1}=\int dt \phi_1(t)\vb{a}^{\dag}(t)$ and $\vb{b}^{\dag}_{\phi_2}=\int dt \phi_2(t)\vb{b}^{\dag}(t)$. With a relative time delay $\tau$ of the arrival time, the beamsplitter input state is
\begin{equation*}
    \ket{\psi}_{in}=\int dt_1 \phi_1(t_1-\tau)\vb{a}^{\dag}(t_1) \int dt_2 \phi_2(t_2)\vb{b}^{\dag}(t_2)\ket{0}.
\end{equation*}
Our beamsplitter is designed to be symmetric \cite{Ou1989} (also see Fig.~S1(c)) so the beamsplitter transformations, with reflection coefficient $\eta$, are
\begin{equation*}
\begin{aligned}
    \vb{a}^{\dag}(t) &\rightarrow i\sqrt{\eta} \, \vb{a}^{\dag}(t) + \sqrt{1-\eta} \, \vb{b}^{\dag}(t), \\
    \vb{b}^{\dag}(t) &\rightarrow \sqrt{1-\eta} \, \vb{a}^{\dag}(t) + i\sqrt{\eta} \, \vb{b}^{\dag}(t).
\end{aligned}
\end{equation*}
The beamsplitter output state is then given by
\begin{equation*}
\begin{aligned}
    \ket{\psi_{out}} &= \int dt_1 \phi_1(t_1-\tau) \int dt_2 \phi_2(t_2) \\
    &\qty[(1-\eta) \, \vb{a}^{\dag}(t_2) \vb{b}^{\dag}(t_1) - \eta \, \vb{a}^{\dag}(t_1) \vb{b}^{\dag}(t_2) + i\sqrt{\eta(1-\eta)} \, [\vb{a}^{\dag}(t_1) \vb{a}^{\dag}(t_2) - \vb{b}^{\dag}(t_1) \vb{b}^{\dag}(t_2)]] \ket{0}.
\end{aligned}
\end{equation*}
The projectors for detection in modes $\vb{a}$ and $\vb{b}$ are $\vu{P}_a = \int dt \,\vb{a}^{\dag}(t) \dyad{0} \vb{a} (t)$ and $\vu{P}_b=\int dt \, \vb{b}^{\dag}(t) \dyad{0} \vb{b} (t)$.
The coincidence probability of detecting one phonon in each mode is
\begin{equation*}
\begin{aligned}
    P_{11}(\tau)&=\Tr(\ket{\psi_{out}} \bra{\psi_{out}}\vu{P}_a \vu{P}_b)=\bra{\psi_{out}} \vu{P}_a \vu{P}_b \ket{\psi_{out}}\\
    &=\bra{0} \int dt_1 \phi_1^*(t_1-\tau) \int dt_2 \phi_2^*(t_2)\\
    &\times\qty[(1-\eta) \vb{a}(t_2)\vb{b}(t_1)-\eta\vb{a}(t_1)\vb{b}(t_2)-i\sqrt{\eta(1-\eta)}[\vb{a}(t_1)\vb{a}(t_2)-\vb{b}(t_1)\vb{b}(t_2)]]\\
    &\times \qty[\int dt_a \vb{a}^{\dag}(t_a) \dyad{0} \vb{a} (t_a) \int dt_b \vb{b}^{\dag}(t_b) \dyad{0} \vb{b} (t_b)]\times \int dt_1' \phi_1(t_1'-\tau) \int dt_2' \phi_2(t_2')\\
    &\times\qty[(1-\eta) \vb{a}^{\dag}(t_2')\vb{b}^{\dag}(t_1')-\eta\vb{a}^{\dag}(t_1')\vb{b}^{\dag}(t_2')+i\sqrt{\eta(1-\eta)}[\vb{a}^{\dag}(t_1')\vb{a}^{\dag}(t_2')-\vb{b}^{\dag}(t_1')\vb{b}^{\dag}(t_2')]] \ket{0}
\end{aligned}
\end{equation*}
By working out the integrals and assuming $\phi_1(t)$ and $\phi_2(t)$ are real and normalized, we arrive at Eq.~(\ref{eq1}). The frequency-domain two-photon coincidence probability is given in Ref. \cite{Branczyk2017}
\begin{equation*}
    P_{11}(\omega_1,\omega_2,\tau) = 1 - 2\eta + 2\eta^2 + (2\eta^2 - 2\eta) \int d\omega_1 \Phi_1^*(\omega_1)\Phi_2(\omega_1) e^{-i\omega_1 \tau} \int d\omega_2 \Phi_2^*(\omega_2)\Phi_1(\omega_2) e^{i\omega_2 \tau}
\end{equation*}
Where $\Phi_{1,2}(\omega)=\int dt \phi_{1,2}(t) e^{-i\omega t}$ are the spectral amplitude functions.

\subsection{Numerical simulations}
To simulate the quantum dynamics of the qubit and phonon wavepacket interactions, we use input-output theory with quantum pulses \cite{Kiilerich2019}. The qubits are modelled as two-level systems without considering higher energy levels. Each qubit is associated with its respective input and output flying-oscillator modes. We constrain each flying oscillator to a single mode with three energy levels, as the maximum excitation number in our system is two phonons (for zero-delay interference of single phonons at the BS). Qubit intrinsic lifetimes, dephasing times, effective phonon lifetimes and phonon release and propagation times are included, with values extracted from measurements. Phonon emission and capture processes are numerically simulated with the open-source Python package QuTiP \cite{Johansson2012}, where we obtain beamsplitter input phonon waveforms and calculate the beamsplitter output phonon states as described in the previous subsection. With the beamsplitter output waveforms and phonon states, we can further simulate the qubit populations during the phonon capture processes. Simulation results are in good agreement with experimental data.

\subsection{Device design and fabrication}
Our device fabrication follows the same process as in Ref.~\cite{Satzinger2018}. The acoustic chip is fabricated with a single layer of 30 nm thick aluminum using a bilayer PMMA liftoff process on a LiNbO$_3$ wafer. The phonon beamsplitter is fabricated using evenly-spaced aluminum fingers. Each finger is 365 nm wide and 150 $\mu$m long, with 143 nm spacing between fingers. The number of fingers is selected experimentally, by fabricating devices where the finger number varied between 12 and 18 with the beamsplitter performance measured at room temperature (See Fig.~S1).  The beamsplitter with $\eta$ closest to $1/2$ was selected for low-temperature measurements; this device has 16 fingers. The unidirectional transducers follow the design in Ref. \cite{Dumur2021}, here however using a single-electrode tapered interdigitated transducer (IDT) design with optimized IDT-mirror spacing. A separate room temperature characterization of devices with identical design shows the UDT directivity is more than 20 dB, with a bandwidth of 150 MHz centered at 3.925 GHz. The qubit circuit, with two qubits, two variable couplers, and the control and readout wiring, is fabricated on a sapphire substrate. Singulated acoustic and qubit circuit dies are flip-chip aligned and attached to one another to form the complete device. Parameters for the device components are displayed in Table~S1.

\subsection{Qubit readout correction}
Two-qubit measurement correction \cite{Bialczak2010} is applied to all the qubit measurement data. We measure both qubits simultaneously using a multiplexed readout pulse. Prior to each experiment, we measure the two-qubit readout visibility matrix, by preparing the two qubits in the fiducial states $\qty\big{\ket{gg}, \ket{ge}, \ket{eg}, \ket{ee}}$ followed by two-qubit readout. The visibility matrix $V$ is defined as the transformation between the measured probability vector and the expected probability vector for the different fiducial states, $P_{meas} = V P_{exp}$. A typical visibility matrix is:
\begin{equation*}
V=\begin{pmatrix}
F_{gg,gg} & F_{gg,ge} & F_{gg,eg} & F_{gg,ee}\\
F_{ge,gg} & F_{ge,ge} & F_{ge,eg} & F_{ge,ee}\\
F_{eg,gg} & F_{eg,ge} & F_{eg,eg} & F_{eg,ee}\\
F_{ee,gg} & F_{ee,ge} & F_{ee,eg} & F_{ee,ee}
\end{pmatrix}=\begin{pmatrix}
0.9806 & 0.0118 & 0.0075 & 0.0001\\
0.0381 & 0.9544 & 0.0003 & 0.0072\\
0.0431 & 0.0006 & 0.9451 & 0.0112\\
0.0018 & 0.042 &  0.0408 & 0.9154
\end{pmatrix}
\end{equation*}
Where $F_{a,b}$ represents the fidelity of preparing two-qubit state to $\ket{a}$ and measuring the two-qubit state on $\ket{b}$. By inverting the visibility matrix we obtain the measurement-corrected two-qubit probability vector $P_{corr} = V^{-1} P_{meas}$. This readout correction affects the visibility of the joint excitation probability dip; the uncorrected visibility of this dip is $\mathcal{V}_{raw}=0.864\pm0.012$, compared to the measurement-corrected value $\mathcal{V}_{corr} = 0.910 \pm 0.013$.


\subsection{Room temperature characterization of phonon beamsplitter}

\begin{figure}[ht!]%
\centering
\includegraphics[width=1\textwidth]{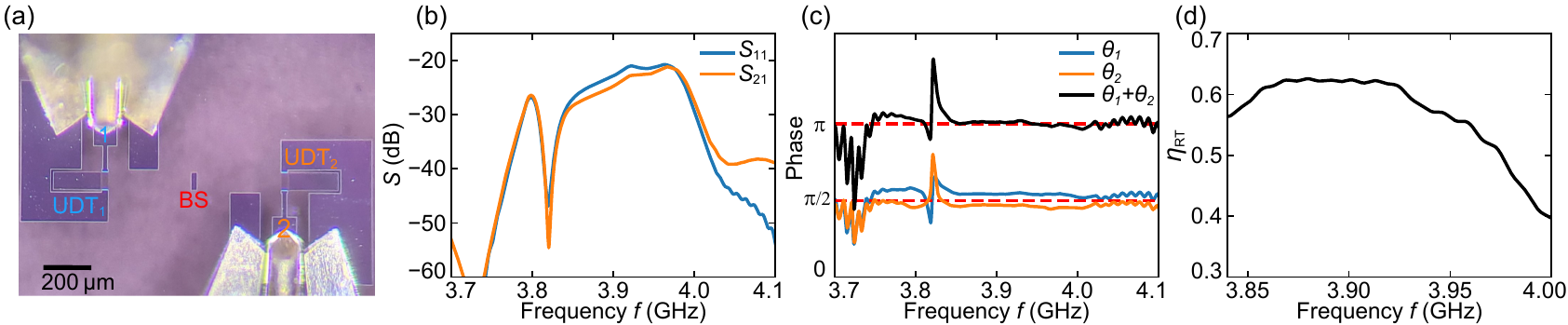}
\captionsetup{labelformat=empty}
\caption{Fig.~S1: \textbf{VNA measurement of phonon beamsplitter.} (a) Optical micrograph of test beamsplitter, fabricated on lithium niobate and measured at room temperature using a microwave probe station and VNA; the device also includes unidirectional transducers (UDTs) to send and detect surface acoustic wave signals. (b) VNA measurement of scattering amplitudes $S_{11}$ and $S_{21}$, representing the $\text{UDT}_1$-BS-$\text{UDT}_2$ performance near the device center design frequency of 3.925 GHz. (c) VNA measurement of BS phase shift from two ports $\theta_1=\angle S_{11}-\angle S_{21}$, $\theta_2=\angle S_{22}-\angle S_{12}$ and the sum of $\theta_1$ and $\theta_2$. Red dashed lines denote the ideal phase shift of $\pi/2$ and the sum of phase shift $\pi$. (d) Beamsplitter reflection coefficient $\eta_{RT}=P_{11}/(P_{11}+P_{21})$, where $P_{11}$ and $P_{21}$ are the powers of scattering amplitudes $S_{11}$ and $S_{21}$.
}\label{figS1}
\end{figure}

Prior to the qubit experiment, we characterized different phonon beamsplitter designs together with UDTs at room temperature with a vector
network analyzer (VNA), using a calibrated microwave probe station. An optical image of a typical test structure of the beamsplitter design is shown in Fig.~S1(a) with VNA ports denoted by 1 and 2. We measure both reflection $S_{11}\approx S_{22}$ and transmission $S_{21}\approx S_{12}$ between UDT$_1$ and UDT$_2$. We display the beamsplitter scattering amplitudes, phase shifts, and room temperature reflection coefficient in Fig.~S1(b-d). We find that even with more than 20 dB of propagation loss at room temperature, the room temperature measurement of the reflectivity $\eta_{RT}=0.612$ at 3.925 GHz agrees well with low-temperature qubit-based measurement of $\eta=0.611 \pm 0.003$ at the same frequency. 

\subsection{Device parameters and addition characterization}
\begin{ruledtabular}
\begin{table}[ht!]
\centering
\begin{tabular}{@{}llll@{}}
Qubit parameters & Qubit 1  & Qubit 2 \\
\hline
Maximum frequency (GHz)    & 4.604   & 4.796\\
Idle frequency (GHz)    & 4.113   & 4.197\\
Anharmonicity (MHz)  &-192 & -191\\
Intrinsic lifetime $T_1$ ($\mu s$)   & 26.7 (3)   & 22.0 (2)\\
Ramsey dephasing time $T_{2,R}$ ($\mu s$)  & 3.0 (1)   & 3.4(2) \\
Spin-echo dephasing time $T_{2,E}$ ($\mu s$)  & 11.2 (2)   & 9.4 (2) \\
Readout resonator frequency (GHz)  & 5.149   & 5.082 \\
$\ket{e}$ state visibility  & 0.959   & 0.957 \\
$\ket{g}$ state visibility  & 0.993   & 0.990 \\
\hline
SAW parameters & Beamsplitter  & Mirror & IDT  \\
\hline
Number of cells (design)  &16&400&28 \\
Aperture ($\mu m$) (design)  &150&150&150 \\
Pitch ($\mu m$) (design) &0.5 & 0.5 & 0.975\\
Metallization ratio (SEM) &0.72&0.72& 0.50 \\
\end{tabular}
\captionsetup{labelformat=empty}
\caption{Table.~S1: \textbf{Summary of device parameters.}}
\end{table}
\end{ruledtabular}

Table S1 gives a summary of the device parameters. Qubit lifetimes, dephasing times, and qubit state visibilities are measured at the qubit idle frequency. $\ket{e}$, $\ket{g}$ state visibilities are the probabilities of measuring $\ket{e}$, $\ket{g}$ states when preparing the qubit on $\ket{e}$, $\ket{g}$ states. One cell in each SAW element is a metal finger with its corresponding gap; the SAW aperture is the width measured perpendicular to SAW emission direction; the pitch is the center-to-center finger distance, and the metallization ratio is the ratio of metal width to gap width. The metallization used here is 30 nm thick aluminum, patterned by electron beam lithography lift-off.

\begin{figure}[ht!]%
\centering
\includegraphics[width=1\textwidth]{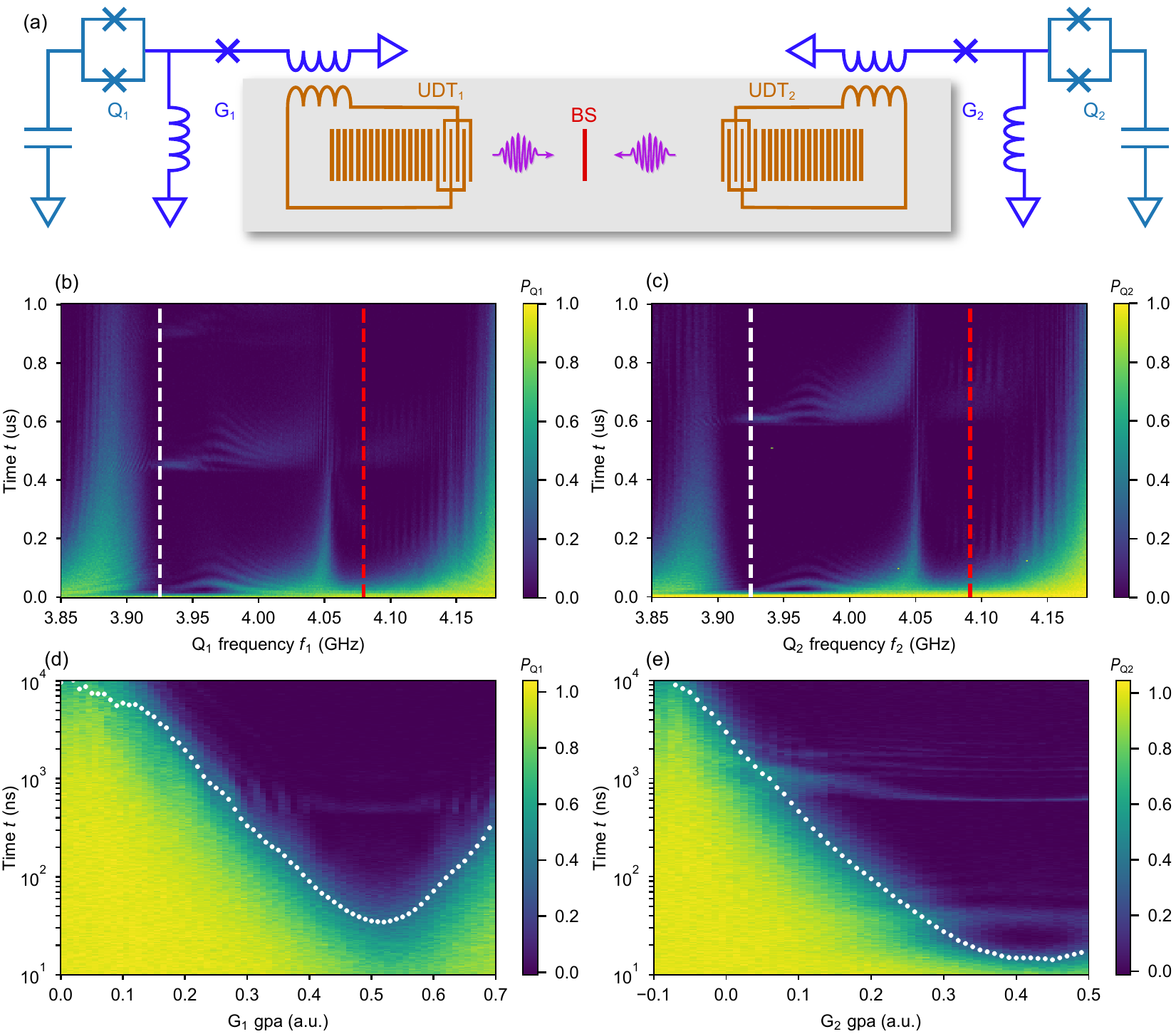}
\captionsetup{labelformat=empty,width=1\linewidth}
\caption{Fig.~S2: \textbf{Circuit diagram and additional device characterization.} (a) Device circuit diagram device representation. (b), (c) Excitation response versus time for $\text{Q}_1$ and $\text{Q}_2$ showing qubit spectrums at acoustic device frequency range. White dashed lines represent the qubit operating frequency of 3.925 GHz. Red dashed lines are thermal dump frequencies, used to thermalize each qubit's excited state population in Fig.~S3. (d), (e) Tunable qubit-UDT coupling strength as a function of coupler $\text{G}_1$ and $\text{G}_2$ control pulse amplitudes (gpa, in arbitrary units) at frequency of 3.925 GHz. White dots are fit $T_1$ times as a function of gpa amplitude. 
}\label{figS2}
\end{figure}

In Fig.~S2(a) we also show the circuit diagram for the device used in the main text. Two qubits $\text{Q}_1$ and $\text{Q}_2$ and two tunable couplers $\text{G}_1$ and $\text{G}_2$ and their respective control and readout lines are patterned on a sapphire die. Each qubit circuit is inductively coupled to a unidirectional transducer, $\text{UDT}_1$ and $\text{UDT}_2$, patterned together with the beamsplitter BS on a separate lithium niobate die. We characterize the lifetime spectrum for $\text{Q}_{1,2}$ by exciting each qubit and monitoring the qubit excited state population $P_\text{Q1,2}$ as a function of time, with the variable coupler set to the same coupling as in the  experiments in the main text. The beamsplitter reflection signals at $\sim 0.45~ \rm \mu s$ and $\sim 0.6~ \rm \mu s$ are visible within the transducer unidirectional band from 3.91 GHz to 4.05 GHz. In the same way, the tunable qubit-UDT coupling strength is measured by monitoring the qubit decay rate as a function of coupler $\text{G}_1$ and $\text{G}_2$ control pulse amplitudes for the initially excited-state qubits set to the operating frequency of 3.925 GHz. We show a wide range of tunable qubit $T_1$ lifetime from $\sim$10 $\mu$s down to $\sim$10 ns.

\subsection{Two-phonon detection at zero relative delay}
\begin{figure}[H]%
\centering
\includegraphics[width=1\textwidth]{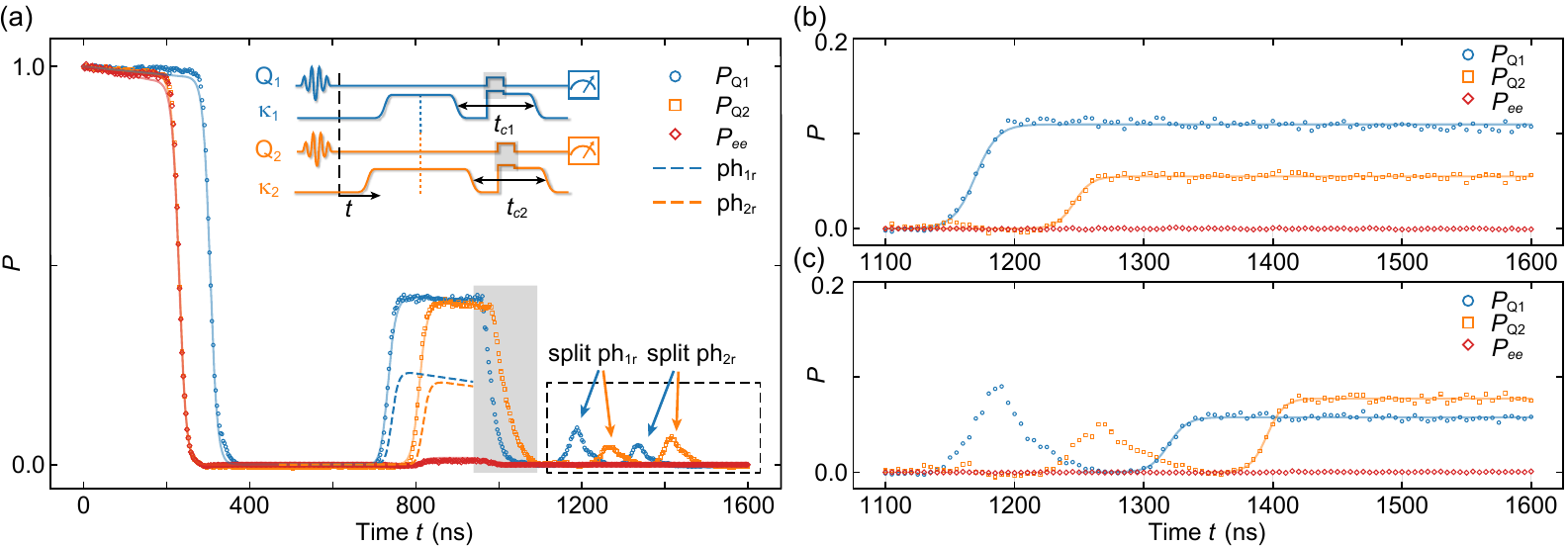}
\captionsetup{labelformat=empty,width=1\linewidth}
\caption{Fig.~S3: \textbf{Two-phonon detection.} (a)  Two-phonon detection at zero relative delay ($\tau=0$) with identical input single phonons. The data (blue and orange points) up to about 900 ns showing this process agrees well with numerical simulations (blue and orange solid lines). Blue and orange dashed lines are simulation results for the phonon populations ph$_{1r}$ and ph$_{2r}$ reflected off Q$_{1,2}$ respectively. After thermally dumping the qubit populations marked in the gray area, four phonon wavepackets split from two reflected phonon packets are denoted by arrows in the black dashed frame. Inset shows the pulse sequence.  (b), (c) Measurements taken following the same sequence as in (a), but with the secondary phonon wavepackets captured by both qubits with adjusted catch time $t_{c1}$ and $t_{c2}$, in (b) turning the coupler off to capture the first two secondary wavepackets split from $\text{ph}_{1r}$ and in (c) to capture the last two wavepackets split from $\text{ph}_{2r}$. We note that our system only allows one measurement of each qubit per repetition; repetitions in which the qubits are measured as in (a) are distinct from repetitions where the qubits are measured as in (b) or (c).
}\label{figS3}
\end{figure}

The two-phonon experiment at zero relative delay, shown in Fig.~3(b) in the main text, is limited in that each qubit can catch a maximum of one phonon at a time. We explore an indirect method to detect the two-phonon state. As shown in Fig.~S3, following emission by the qubits, controlled via the coupling rates $\kappa_{1,2}(t)$, interference of the emitted phonons at the beamsplitter results in a two-phonon output state from the BS. Ideally this state is $(\ket{20}+\ket{02})/\sqrt{2}$, written in the basis $\ket{\text{ph}_{1} \text{ph}_{2}}$, where $\text{ph}_{1,2}$ denote phonon numbers in the output phonon channels directed towards $\text{Q}_{1,2}$. The qubits can at most catch one phonon from this output, in the ideal process $\ket{g20g}+\ket{g02g} \rightarrow \ket{e10g}+\ket{g01e}$, with states written as $\ket{\text{Q}_1 \text{ph}_1 \text{ph}_2 \text{Q}_2}$ and we ignore normalization. The data showing this process (blue and orange points) up to about 900 ns agree well with numerical simulations including phonon decay \cite{Johansson2012} (blue and orange solid lines). The blue and orange dashed lines are simulation results for the phonon populations ph$_{1r}$ and ph$_{2r}$ reflected off Q$_{1,2}$, respectively, directed towards the BS and not absorbed by the qubits, therefore not detected in the experiment. These phonons propagate and interact with the BS, where the off-center location of the BS prevents further phonon-phonon interference, splitting these secondary phonons into a total of four wavepackets directed at the qubits. In one variation of this experiment, following interaction with the initial two-phonon state, the qubits are reset to $|g\rangle$ by tuning the qubits to the thermal dump frequency roughly 100~MHz away from 3.925 GHz (red dashed line in Fig.~S2) and turning on their couplers, thermally dumping any excitation into acoustic reservoirs at frequencies irrelevant to the experiment (gray shaded areas in pulse sequence and data). The qubits relax to their ground states during the thermal dump. Each qubit's coupler is then adjusted and the qubit frequency is reset to the phonon frequency, allowing the qubits to absorb the remaining four phonon wavepackets, as denoted by arrows in the black dashed frame. As In Fig.~S3(a) the couplers are left on in this process, the qubits then re-emit these excitations and return to their ground states. In Fig.~S3(b) and (c), we instead capture the secondary phonon wavepackets, by using the qubits with adjusted catch times $t_{c1}$ and $t_{c2}$, adjusted in B to capture the first pair of secondary wavepackets resulting from the interaction of ph$~{1r}$ with the beamsplitter and in C to capture the last pair of secondary wavepackets. In these alternative measurements, $P_{ee}$ remains close to zero as expected after the capture, as the phonon wavepackets are split from the same reflected phonon.

\subsection{Two-phonon interference experiment with different phonon wavepacket widths}
\begin{figure}[ht!]%
\centering
\includegraphics[width=1\textwidth]{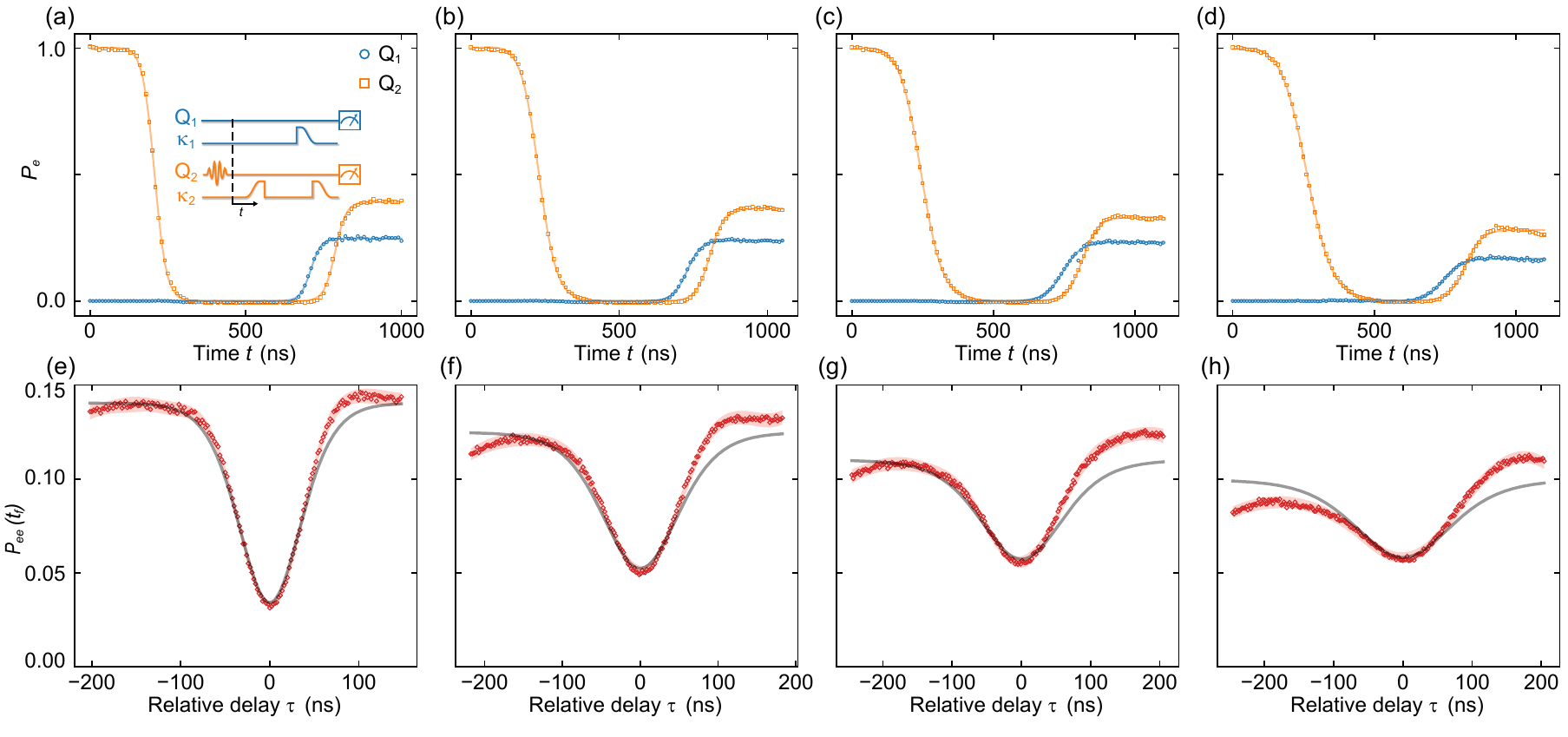}
\captionsetup{labelformat=empty}
\caption{Fig.~S4: \textbf{Two-phonon interference experiment with different phonon wavepacket widths.} (a-d) Tune-up for emission measurements with different wavepacket widths, emitted from $\text{Q}_2$. The fit phonon packet widths are $\sigma_2$ = 18.7 ns, 28.9 ns, 36.4 ns and 43.3 ns. (e-h) Fixing Q$_1$'s wavepacket width to $\sigma_1$ = 8.8 ns, and performing the two-phonon interference experiment with packet widths $\sigma_2$ for Q$_2$, using the calibration from panels (a-d).
}\label{figS4}
\end{figure}

In Fig.~4(b) of the main text, we show the dependence of the HOM dip visibility on the relative phonon wavepacket width. In Fig.~S4, we display the two-phonon interference results for each data point in Fig.~4(b). In Fig.~S4(a-d), we show the response for different wavepacket widths $\sigma_2$ released from $\text{Q}_2$. In Fig.~S4(e-h), we perform the two-phonon interference experiment with different wavepacket widths $\sigma_2$ = 18.7 ns, 28.9 ns, 36.4 ns and 43.3 ns, while keeping Q$_1$'s wavepacket width at $\sigma_1$ = 8.8 ns. The HOM dip visibility becomes smaller as difference between the two phonon wavepackets increases. We note the catch efficiency becomes lower for larger phonon wavepacket width in panels (a-d), because the entire wavepacket spatial length is running into the limit set by the distance between the UDTs and the BS. The lower catch efficiency for $\text{Q}_2$ further increases the asymmetry of the HOM dip in panels (e-h).

\subsection{Two-phonon interference with arbitrary phonon wavepacket shape}
\begin{figure}[ht!]%
\centering
\includegraphics[width=1\textwidth]{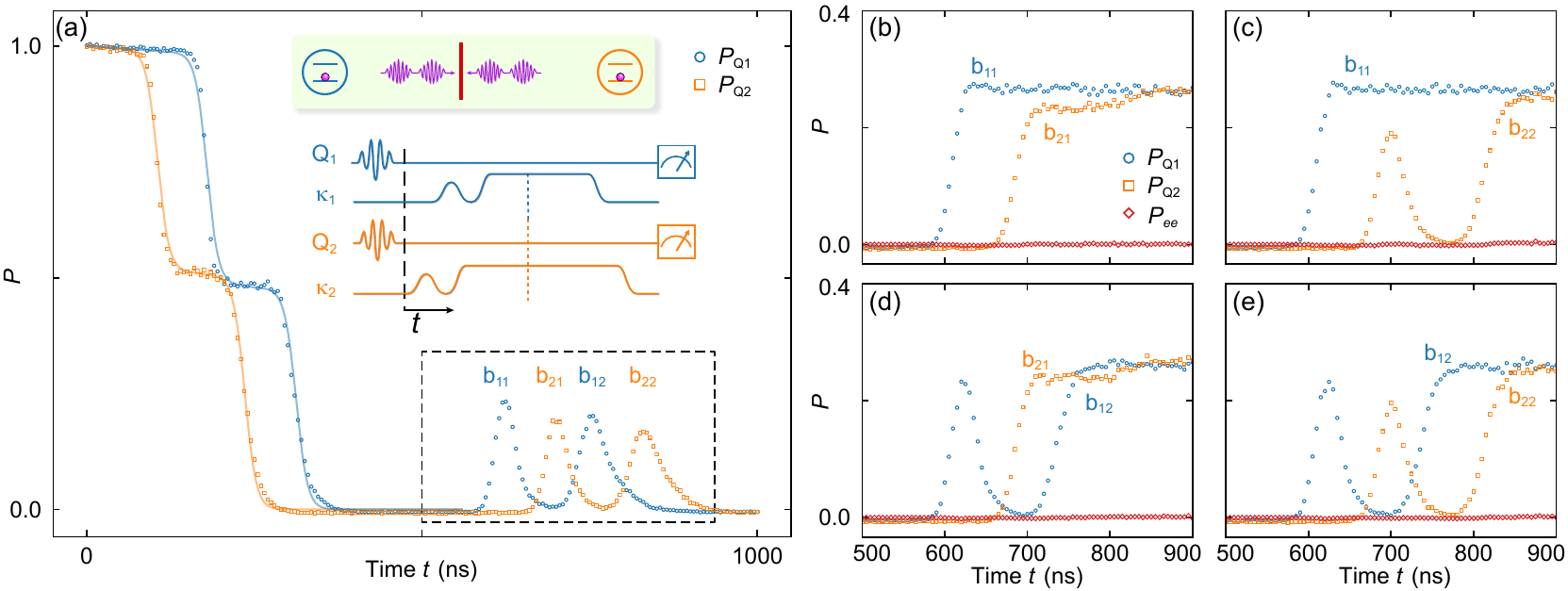}
\captionsetup{labelformat=empty}
\caption{Fig.~S5: \textbf{Two-phonon interference experiment with time-bin encoded phonons.} (a) For each qubit, its phonon is released as two half-phonons in two separate time bins, and each pair of time-bin encoded half-phonons arrives at the beamsplitter with zero relative delay. Four events (within the dashed frame) are detected, with both variable couplers left on. Inset shows the schematic and pulse sequence. (b-e) Absorption of the beamsplitter output waveforms, timed to catch each event in panel (a), showing pairwise coincidence detection.
}\label{figS5}
\end{figure}

As mentioned in the main text, the functional dependence of two-phonon interference can be extended to arbitrary waveforms using our phonon pulse shaping capabilities. In Fig.~S5, we show an example, where each of the two phonons in the two-phonon interference experiment is distributed in two time bins. This is achieved by releasing half of the qubit excited state population in the first time bin, with the remaining population released in a later time bin. As shown in the schematic in panel (a), each pair of time-bin encoded half-phonons arrives at the beamsplitter with zero relative delay and the output of the beam splitter is detected with the couplers left on. A total of four events are detected by the two qubits, denoted by $b_{ij}$, where $b_{ij}$ represents the $j$th time-bin wavepacket captured by qubit $\text{Q}_i$. We show pairwise coincidence detection in panel (b-e): $b_{11}$ and $b_{21}$, $b_{11}$ and $b_{22}$, $b_{12}$ and $b_{21}$, $b_{12}$ and $b_{22}$, respectively. We find the joint excitation probability $P_{ee}<0.006$ for all four cases, which indicates two-phonon interference still occurs with indistinguishable single phonons distributed in two time bins.

\end{document}